\magnification=1200
\def\nonumfirst{\nopagenumbers
                \footline={\ifnum\count0=1\hfill
                           \else\hfill\folio\hfill
                           \fi}}
\nonumfirst
\magnification=1200
\def\singlespace{\baselineskip 12 pt}

\def\oneandahalfspace{\baselineskip 18pt}
\def\blankline{\vskip 12 pt\noindent}

\def\secto#1\endsecto{\vskip 20pt {\bf #1}\vskip 7pt\nobreak}
\global\newcount\refno \global\refno=1
\newwrite\rfile
\def\ref#1#2{\hbox{[\hskip 2pt\the\refno\hskip 2pt]}\nref#1{#2}}
\def\nref#1#2{\xdef#1{\hbox{[\hskip 2pt\the\refno\hskip 2pt]}}%
\ifnum\refno=1\immediate\openout\rfile=refs.tmp\fi%
\immediate\write\rfile{\noexpand\item{\noexpand#1\ }#2.}%
\global\advance\refno by1}
\def\semi{;\hfil\noexpand\break}
\def\demi{:\hfil\noexpand\break}

\newwrite\efile \let\firsteqn=T
\def\writeqno#1%
{\if T\firsteqn \immediate\openout\efile=eqns.tmp\global\let\firsteqn=F\fi%
\immediate\write\efile{#1 \string#1}\global\advance\meqno by1}

\def\eqnn#1{\xdef #1{(\the\secno.\the\meqno)}\writeqno#1}
\def\eqna#1{\xdef #1##1{(\the\secno.\the\meqno##1)}\writeqno{#1{}}}

\def\eqn#1#2{\xdef #1{(\the\secno.\the\meqno)}%
$$#2\eqno(\the\secno.\the\meqno)$$\writeqno#1}
\def\nobreak{\penalty1000}
\def\titl#1\endtitl{\par\vfil
     \vbox to 2in {}{\bf #1}\nobreak}
\def\titol#1\endtitol{\par\vfil
     \par\vbox to 1in {}{\bf #1}\par\vskip 1in\nobreak}
\def\tit#1\endtit{
     \vbox to 0.5in {}{\bf #1}\nobreak}

%
%
%
%
%
%
%
\def\lspace{\ifx\answ\bigans{}\else\qquad\fi}
\def\lbspace{\ifx\answ\bigans{}\else\hskip-.2in\fi} 
%
%
%

 \def\CZ{{\cal Z}}
%
%
%
%

%

\def\bar#1{\overline{#1}}

\def\bra#1{\left\langle #1\right|}
\def\ket#1{\left| #1\right\rangle}
\def\abs#1{\left| #1\right|}

%
%
\def\frac#1#2{{\textstyle{#1\over #2}}} 
%
%
%
%

\def\Im{\mathop{\rm Im}}
\def\kok{ {\rm K}^ 0 - \overline{\rm K}{}^0 }

\def\ko{ {\rm K}^ 0 }
\def\kob{ \overline{\rm K}{}^0 }
\def\kl{ {\rm K_{L}} }
\def\ks{ {\rm K_{S}} }

\def\gl{ {\it \Gamma}_{\rm L} }

%
%
%
\def\ltap{\ \raise.3ex\hbox{$<$\kern-.75em\lower1ex\hbox{$\sim$}}\ }
\def\gtap{\ \raise.3ex\hbox{$>$\kern-.75em\lower1ex\hbox{$\sim$}}\ }
\def\gl{\ \raise.5ex\hbox{$>$}\kern-.8em\lower.5ex\hbox{$<$}\ }
\def\roughly#1{\raise.3ex\hbox{$#1$\kern-.75em\lower1ex\hbox{$\sim$}}}
%
%

%
\def\[{\left[}
\def\]{\right]}
\def\({\left(}
\def\){\right)}
%
%

%
\textfont2=\tensy \scriptfont2=\sevensy \scriptscriptfont2=\fivesy
\def\cal{\fam2}
\def\Ascr{{\cal A}}
\def\Bscr{{\cal B}}
\def\Iscr{{\cal I}}
\def\Pscr{{\cal P}}
\def\Qscr{{\cal Q}}

\def\Hscr{{\cal H}}
\def\Gscr{{\cal G}}

\def\Uscr{{\cal U}}
\def\Vscr{{\cal V}}
\def\Sscr{{\cal S}}
\def\hA{{\widehat A}}

\def\st{\scriptstyle}

\def\pmb#1{\setbox0=\hbox{$#1$}%
  \kern-.025em\copy0\kern-\wd0
  \kern.05em\copy0\kern-\wd0
  \kern-.025em\raise.0433em\box0}
\def\pmbs#1{\setbox0=\hbox{$\st #1$}%
  \kern-.0175em\copy0\kern-\wd0
  \kern.035em\copy0\kern-\wd0
  \kern-.0175em\raise.0303em\box0}

\def\bfs#1{\hbox to .0035in{$\st#1$\hss}\hbox to .0035in{$\st#1$\hss}\st#1}

\def\bfDelta{\pmb{\Delta}}
\def\bfPi{\pmb{\Pi}}
\def\bfPsi{\pmb{\Psi}}
\def\bfLambda{\pmb{\Lambda}}
\def\bfGamma{\pmb{\Gamma}}
\def\bfDeltap{\pmb{\Delta^\prime}}
\def\Ttr{\pmb  \tau}

\def\bfI{\hbox{\bf I}}

\def\bfR{\hbox{\bf R}}

%

\def\bra#1{\langle #1 \vert }
\def\ket#1{\vert #1 \rangle }

%
%
\global\newcount\meqno \global\meqno=1
\newwrite\efile \let\firsteqn=T
\def\writeqno#1%
{\if T\firsteqn \immediate\openout\efile=eqns.tmp\global\let\firsteqn=F\fi%
\immediate\write\efile{#1 \string#1}\global\advance\meqno by1}

\def\eqqn#1#2{\xdef #1{(\the\meqno)}%
$$#2\eqno(\the\meqno)$$\writeqno#1}
\font\medf=cmb10 scaled \magstep3
%
\vsize=25 truecm
\hsize=16 truecm
\voffset=-0.8 truecm
%
\singlespace
\parskip 6truept
\parindent 20truept
\vbox{ {\rightline{\bf IFUM--FT 600/97}}
       {\rightline{\bf UNIBAS--MATH 9/97}}
}
\hyphenation{ex-pe-ri-men-tal}
\hyphenation{va-cu-um}
\vskip 4truecm
\centerline{\medf 
                The Quantum Field Theory of the Kaon Oscillations}
\vskip 2truecm
\vskip 33truept
\centerline{D. Cocolicchio$^{(1,2)}$ and M. Viggiano$^{(1)}$}
\vskip 20truept
\vbox{
\centerline{\it $^{1)}$Dipartimento di Matematica,
Univ. Basilicata, Potenza, Italy}
\vskip 5truept
\centerline{\it Via N. Sauro 85, 85100 Potenza, Italy} }
\vskip 15truept
\vbox{
\centerline{\it $^{2)}$Istituto Nazionale di Fisica Nucleare,
                     Sezione di Milano, Italy}
\vskip 5truept
\centerline{\it Via G. Celoria 16, 20133 Milano, Italy} }
\vskip 3truecm
\centerline{\it ABSTRACT}
\vskip 15truept
\singlespace
\vskip 1truecm
\noindent
{\sl We consider the covariant formulation of the kaon mixing in the 
context of the propagator method. It results important to check the 
possibility of a sizable effect in the vacuum regeneration of kaons.
We discuss all those terms which may give relevant contributions to 
modify the exponential decay law of the Wigner--Weisskopf narrow width 
approximation. Moreover, we examine the characteristic structure of 
the complex singularities of the matrix propagator and we provide a 
generalized form of the Bell-Steinberger unitarity sum rule.
}
\vskip 1truecm \noindent \singlespace
\vbox{
      {\leftline{PACS numbers: 11.30.Er, 12.90.+b, 13.20.Eb. }}
      {\leftline{\it Keywords: CP Violation, Propagator Method, Oscillations.}}
      }
\vfill\eject
\vsize=24 truecm
\hsize=16 truecm
\baselineskip 18 truept
\parindent=1cm
\parskip=8pt
\oneandahalfspace
%
%
\phantom{.}
\blankline
\blankline
\leftline {\bf I. Introduction}
\blankline
\noindent
The origin of $CP$ violation is still not explained, although the Standard
Model of the electroweak interactions can accomodate a complex violating phase
in the quark mixing matrix
\ref\Pavia{D. Cocolicchio,
``{\it CP--asymmetries in $B$ decays}'',
Proc. 
{\it Advanced Study Conference on Heavy Flavours},
ed. G. Bellini et al. (Ed. Frontieres, 1993), p. 367}.
However, the eventuality that $CP$ violation originates from some effects
at a much higher energy scale is not excluded. Indeed, the high sensitivity
of the $\kok$, makes it a testing ground of even more
speculative proposals like $CPT$-violation
\ref\CPT{V. A. Kostelecky and R. Potting, Phys. Rev. {\bf D51} (1995) 3923} \ 
and deviations from the conventional Quantum Mechanics
\ref\QMCPT{
F. Benatti and R. Floreanini, Phys. Lett. {\bf B389} (1996) 100 and 
refs. therein}.
In view of the planned high precision experiments in kaon physics, it
is then important to reconsider the fundamental aspects of the 
space-time evolution of the $\kok$ system also beyond the generalized
single pole or Lee-Oehme-Yang (LOY) approximation
\ref\LOY{T. D. Lee, R. Oehme and C. N. Yang, Phys. Rev. {\bf 106} 
(1957) 340; P. K. Kabir, {\it The CP Puzzle} (Academic Press,
New York, 1968), Appendix A}.
\hfill \break
Physically, it is worthwhile to examine the possibility of a
sizable effect in the vacuum regeneration of kaons
\ref\KAK{L. A. Khalfin, preprint Univ. Texas at Austin CPT-Report DOE-ER
40200-211 (1990); C. B. Chiu and E. C. G. Sudarshan, Phys. Rev. {\bf D42}
(1990) 3731; L. A. Khalfin, ``{\it Comments on a recent paper on 
neutral kaon decay}'', 1991 preprint unpublished; 
K. Urbanowski, Int. J. Mod. Phys. 
{\bf A10} (1995) 1151; Yu. A. Azimov, Phys. Atom. Nucl.
{\bf 59} (1996) 856;
T. Mochizuki, N. Hashimoto, A. Shinbori and S.Y
Tsai, ``{\it Remarks on Theoretical Frameworks Describing the Neutral 
Kaon System}", preprint NUP-A-97-14 (1997)}.
We cannot neglect another complaint of the LOY formalism which seems
to exclude neutral mesons systems from the realm of its sensible applications.
In fact, its assumption that the mixing of elementary particles is 
independent of the momentum, appears too drastic. The inclusion of the 
$q^2$ dependence, usually thought to be small, could have unexpectedly 
large effects mainly in the resonant $CP$-violating processes
\ref\Skaon{D. Cocolicchio, ``{\it The Scattering Theory of $CP$ 
Violation}'', preprint Sezione INFN Milano IFUM-FT 556/96}.
The usual method of the LOY approximation neglects these effects.
Another severe limitation in the application of this approximation 
consists in the fact that it rests completely inappropriate to 
implement the notion of the rest frame for an oscillating unstable 
composite system. In this case, there is a subtle point worth noting 
here, referring to the phase ambiguities that arise in non 
relativistic theories when the superposition of states with different 
mass and momentum are described in a Galilean invariant form
\Skaon.
Indeed, an explicit relativistic description of the neutral meson
systems seems to be required. It is important to stress that if we 
consider the problem in the formalism of the group representations, 
we may find difficulties
to describe unstable particles on the same footing of the stable 
ones because non-unitary representations of the Poincar\'e group
are requested by a fundamental complex mass~\Skaon.

\noindent
In this paper, we consider the covariant formulation of the kaon 
mixing in the context of the propagator method.
Similar works were done along this line
\ref\Kpropag{
R. G. Sachs, Ann. Phys. {\bf 22} (1963) 239;
S. Coleman and H. Schnitzer, Phys. Rev. {\bf 134} (1964) B863;
O. Nachtmann, Acta Phys. Austr. Suppl. {\bf 6} (1969) 485;
D. Sudarsky, E. Fishbach, C. Talmadge, S. H. Aronson and H. Y. Cheng,
Ann. Phys. {\bf 207} (1991) 103;
J. Liu and G. Segre, Phys. Rev. {\bf D49} (1994) 1342}, 
but here we do not use a specific dynamical model and general 
$q^2$-dependent relations are considered. In fact, the structure of 
the propagator is considered only from the general assumptions of 
Lorentz invariance and causality. Furthermore, we discuss all those 
terms which may give significant contributions and we do not restrict 
to the pole approximation which gives the dominant exponential law in
the time evolution. Moreover, we consider the analytical continuation of 
the Fourier transformed propagator in the second Riemann sheet and we 
analyze the characteristic structure of the complex singularities.
Apart the previous theoretical warnings, it will be clear that there 
are some advantages to work with the propagator method rather than in 
the LOY approximation. Indeed, it can be easily seen that the main 
results are equivalent.

\blankline
\leftline {\bf II. The Shortcomings of the Pole Approximation in the
Kaon System}
\blankline
\noindent     
The phenomenological description of the evolution of a $\ko$ meson
consists in the application of a generalized Wigner--Weisskopf narrow 
width approximation in the $\ko$, $\kob$ subspace. Usually, one 
restricts to a single pole or LOY approximation which 
is supplemented by the unitarity sum rule of Bell and Steinberger
\ref\BS{
J. S. Bell and J. Steinberger, Proceedings Oxford Int. 
Conf. on Elementary Particles 1965, ed. R. G. Moorehouse et al.
(Rutherford HEP Lab., Chilton, Didcot, Berkshire, England, 1966) p. 
195}, connecting the two dimensional subspace with the space of all
final states for the decaying system. In this method, the neutral
$K$-meson system is described by a scattering theory 
with the Hamiltonian $H = H_0 + H_{int}$, where $H_0$ contains the strong interactions
under which the $\ko$ and $\kob$ mesons appear as stable particles, and
$H_{int}$ induces their decay into the continuous spectrum of $H_0$.
Let $\Hscr = \Hscr _K \oplus \Hscr _F$ be the Hilbert space of the
neutral $K$ mesons together all their decay products and $\Pscr$
the projection into the two-dimensional subspace $\Hscr _K$ spanned
by $\ko$, $\kob$ (or by $K_{1,2} = {1 \over \sqrt 2}(K_0 \pm \overline{K}
{}^0)$) and $\Qscr = \Iscr - \Pscr$ the projection on the continuous
final states part $\Hscr _F$ of the spectrum $H_0$. Then $H_{int}$ induces
the decay of the $K$-mesons only if $\Qscr H_{int} \Pscr \ne 0$, so that
the evolution of the whole system is given by the dynamical semigroup
relation:
\eqqn\rela{
\Uscr (t) = \exp [-iHt] \quad .}
It is well-known, however, that the projection of this time evolution
into a subspace cannot, in general, have the semigroup property, if
$H$ is not Hermitian. Therefore, the total evolution of Eq.{\rela}
does not conserve the subspace $\Hscr _K = \Pscr \Hscr$ of the Hilbert
space $\Hscr$. In fact, the evolution of the $K$-meson system is governed
by
\eqqn\gov{
\Uscr^ {\prime}(t) = \Pscr \Uscr (t) \Pscr}
where $\Pscr$ does not commute with $H$ and then $\Uscr^ {\prime}(t)$
does not satisfy the semigroup law
\eqqn\law{
\Uscr^ {\prime}(t_1) \Uscr^ {\prime}(t_2) \ne \Uscr^ {\prime}(t_1 + t_2)
\quad .}
On the other hand, the Wigner--Weisskopf (WW) approximation with constant
(not Hermitian) Hamiltonian assures the validity of the semigroup
property. On the other side, the assumption of constant decay 
rates as they arise in the unitary sum rule is also not justifiable
if the time-reversal $T$ invariance
is not a symmetry of the underlying Hamiltonian
\ref\kapil{B.G. Kenny and R.G. Sachs, Phys. Rev. {\bf D8}(1973)1605;
P.K. Kabir and A. Pilaftsis, Phys. Rev. {\bf A53}(1996)66;
H. Yamamoto, Phys. Rev. Lett. {\bf 79}(1997)2402}.
In fact, in this case, the condition of reciprocity is not properly satisfied.
Relaxing the WW pole approximation,
we shall now obtain an approximate form for $\Uscr^ {\prime}(t)$
which enable us to derive the unitary sum rule in a generalized form.

\noindent
In general,
$\Uscr (t)$ can be represented as the Laplace transform of the resolvent
$\Gscr (s) = [s\Iscr - H]^{-1}$
\eqqn\res{
\Uscr (t) = {1 \over {2 \pi i}} \oint \Gscr (s) \exp(-ist) \, ds}
where the integration path is around the spectrum of $H$. Hence
$\Uscr^ {\prime}(t)$ can be expressed in terms of the analytic properties
of the reduced resolvent (the propagator in the $K$ meson space)
\eqqn\red{
\Gscr^ {\prime} (s) = \Pscr \Gscr (s) \Pscr }
in the form
\eqqn\form{
\Uscr^ {\prime} (t) = {1 \over {2 \pi i}} \oint \Gscr^ {\prime} (s)
\exp(-ist) \, ds }
where $\Gscr$ and $\Gscr^ \prime$ are $2 \times 2$ matrices.
The resolvent $\Gscr (s)$ of
the total Hamiltonian $H$ can be assumed analytic on the entire complex
plane except for the spectrum of $H$.
Since the discontinuity of $\Gscr (s)$
is unbounded across the cut along the spectrum of $H$, the cut forms a
natural boundary and the resolvent cannot be analytically continued across.
For a suitable $H_{int}$, however, the discontinuity of the reduced resolvent
$\Gscr^ \prime (s)$ can be regular on some open set belonging to the
spectrum of $H$ and there may then exist an analytic continuation of
$\Gscr^ \prime (s)$. In a model in which $\Qscr H_{int} \Qscr = 0$ (no final
state interactions), and for sufficiently weak coupling, we may assume that
the degenerate eigenvalues in the discrete spectrum of $H_0$ appear as
two poles in the second sheet of the reduced total resolvent
$\Gscr^ \prime$ which are not-degenerate due to the different phase space 
strength of coupling to the decay channels. We further assume that the
rank of $\Gscr^ \prime (s)$ remains two in its domain of regularity, and
hence $\Gscr^ \prime (s)$ admits an inverse. The inverse of
$\Gscr^ \prime (s) \equiv \bfDelta^\prime(s) $ restricted to the subspace 
$\Pscr \Hscr = \Hscr_K$ is
$(\bfDelta^ \prime (s))^{-1}$ so that
\eqqn\that{
\Gscr^ \prime (s) \bfDelta^ {\prime {-1}} (s) = \bfDelta^ {\prime {-1}} (s)
\Gscr^ \prime (s) = \Pscr \quad .}

In order to give our notation some physical content, we may consider 
the specific problem of the kaon mixing, although it should be clear 
that the method is general.
In absence of weak interactions, $\ko$ and $\kob$ are eigenstates 
of the strong interactions and form a degenerate 
particle--antiparticle pair in flavour state, with a common mass 
$m_\circ$ (whatever we assume $CPT$ invariance). 
When higher order weak interactions are 
introduced, transitions are induced between $\ko$ and $\kob$. Thus, 
mixing prohibits the $\ko$, $\kob$ scalar mesons
from propagating independently of each other.
Including the effects of the weak interactions, the full 
dressed matrix propagator is given by
\eqqn\eIIIst{
\bfDeltap (q^2) = \left[ q^2 \bfI - \bfLambda (q^2)\right]^{-1}\; ,
}
where the effective $\bfLambda$ matrix consists in the following
sum of the bare square-mass matrix and
the proper self-energy $\bfPi$ contributions
\eqqn\eIIIo{
\bfLambda (q^2) = \left[ m^2_\circ \bfI + \bfPi (q^2) \right] \, ,
}
in a representation referring to the set of the
factorized constituents $\ko$, $\kob$.
As a consequence of $CPT$ invariance, the diagonal matrix elements are 
equal, whereas the off-diagonal elements $\Delta^\prime_{ij}$
are equal only in the case the interactions are all $CP$ invariant.
Anyway, any invariance of an underlying theory will reflect itself in an 
invariance of the propagator and then also of the square mass matrix 
$\bfLambda$. 
Dropping the superscript prime, the time evolution of the flavour states
\eqqn\eIIu{
\left( \matrix{ \vert \ko (t)  \rangle \cr 
                    \cr
                \vert \kob (t) \rangle \cr}\right) 
=  \Uscr (t) 
\left( \matrix{ \vert \ko \rangle \cr 
                 \cr
          \vert \kob \rangle \cr} \right) \quad ,
}
and similarly for the mass right--eigenstates
\blankline
\eqqn\eIId{
\left( \matrix{ \vert K_S (t) \rangle \cr 
                 \cr
                \vert K_L (t) \rangle \cr}\right) 
=  \Vscr (t) 
\left( \matrix{ \vert K_S \rangle \cr 
                 \cr
                \vert K_L \rangle \cr}\right) \quad 
}
are governed by the evolution matrices $\Uscr$ 
and $\Vscr$ which are related by the following similarity transformation
\eqqn\eIIq{
\Uscr=\bfR^{-1} \Vscr \bfR \quad .
}
The transformation between the physical and flavour
bases are given by
\eqqn\eIIo{
\left( \matrix{ \vert K_S \rangle \cr 
            \cr
          \vert K_L \rangle \cr}\right) =
\bfR
\left( \matrix{ \vert \ko \rangle \cr 
     \cr     
     \vert \kob \rangle \cr}\right) \; ,
}
where $\bfR$ is usually parameterized
according to the following relations
\eqqn\eIIse{
\bfR =  
 {1\over {\sqrt{2(1+\vert\epsilon\vert^2)}}} 
\left(
\matrix{
(1+\epsilon) & -(1-\epsilon) \cr
(1+\epsilon) &  (1-\epsilon) \cr}
\right) =
{1\over{\sqrt{1+\vert\eta\vert^2}}}
\left( \matrix{ 1 & \eta \cr
         1 & -\eta \cr}
\right)
= \left(
\matrix{
p & -q \cr
p & q \cr}
\right)
}
where the normalization factor $(\vert p \vert ^2 + \vert q \vert ^2)^
{-{1 \over 2}}$ is assumed here unity.
\noindent
We remember that the phases of $p$, $q$ may be altered by redefining 
the phases of the $K$ states, so that both $p$ and $q$ are not 
measurable quantities, whereas the overlap between $\ks$ and $\kl$
\eqqn\eIInn{
\langle K_S \vert K_L \rangle = 
{ {2 {\rm Re} \, \epsilon}
     \over
  {1 + |\epsilon|^2} } = 
 { {1 -\vert\eta\vert^2}
      \over
   {1+\vert\eta\vert^2} } 
\quad ,
}
is independent of any phase convention at the same strength of
the magnitude of the variable
\eqqn\eIInnb{
\eta = -{q\over p} = -{{1-\epsilon}\over{1+\epsilon}} 
}
Assuming $\Delta\Sscr = \Delta \Qscr$ rule conserved,
this last quantity is directly connected to the amount of the kaon semileptonic
charge rate
\eqqn\esl{
A_{SL} =
{{\Gamma(K_L\rightarrow \ell^+\nu X)-\Gamma(K_L\rightarrow \ell^-\nu X)}
\over
{\Gamma(K_L\rightarrow \ell^+\nu X) +\Gamma(K_L\rightarrow \ell^-\nu X)}}
=
{{1-|\eta|^2}\over{1+|\eta|^2}} .
}
Nevertheless,
in order to describe the kaon system in terms of uncoupled channels, 
we need to diagonalize $\bfDelta ^\prime$. 
The fact that $\bfLambda$ is, in general, momentum dependent does not
introduce any additional complications, in practice, since $s=q^2$ is
always fixed by the on-shell condition of the initial particles. Anyway,
the resulting eigen-physical fields are those with a definite propagation 
behaviour. 

\noindent
Neglecting the prime superscript,
the regularized inverse propagator can be rewritten as
\eqqn\eIIIduno{
\Delta^{-1}_{ij} = \left[ s \bfI - \bfLambda \right] =
R^{-1}_{i\alpha} \, \Delta^{-1}_{\alpha\beta} \,
R_{\beta j}
}
where
\eqqn\eIIIddue{
\Delta^{-1}_{\alpha\beta}(s) =\left[ s \delta_{\alpha\beta} - 
N_{\alpha\beta} \right]
}
and
\eqqn\eIIIdtre{
N_{\alpha\beta} = {\left[\bfR \bfLambda 
\bfR^{-1}\right]}_{\alpha\beta}\quad .
}
Here, according to a standard prescription \Skaon, the latin indices denote
$\ko$, $\kob$ and the greek letters denote $K_S$, $K_L$, respectively.
Although, in general,
$\bfPi$ (and hence $\bfLambda$), is momentum dependent, in principle
the matrix propagator
can be brought into a diagonal form
\eqqn\eIIIn{
\Delta_{\alpha\beta} (q^2) = \left( \matrix{ \Delta_S 
(q^2) & 0 \cr
0 & \Delta_L (q^2) \cr} \right) 
}
through a $q^2$-dependent transformation. 
It is worth noting that $\bfLambda (q^2) $ shares all the 
properties of the effective Hamiltonian in the description of the 
kaon system. In particular, $CPT$ invariance requires that 
$\Lambda_{11} =\Lambda_{22}$ and $CP$ invariance prescribes the 
equality of the off-diagonal elements $\Lambda_{12} =\Lambda_{21}$.
Thus, the $\kok$ 
mixing gives rise to $CP$ violation through the effective mass-squared 
matrix $\bfLambda (q^2)$. The basic parameter which characterizes the 
indirect $CP$ violation induced by the mixing in the kaon system is 
then given by
\eqqn\eIIIxas{
\eta = {{-q}\over p} = {{-(1-\epsilon)}\over{(1+\epsilon)}} =
\sqrt{ {\Lambda_{21}}\over{\Lambda_{12}} }
}
which is a rephasing invariant quantity and hence physically 
meaningful.

\noindent
The LOY approximation is equivalent to assume the following weak 
coupling
\eqqn\eIIIxiii{
\Pi_{ij} (q^2) \simeq \Pi_{ij} (m^2_\circ) \; .
}
In this case,
\eqqn\eIIIxiv{
\bfLambda = \left[ m^2_\circ \bfI + \bfPi(m^2_\circ)\right]
}
can be diagonalized by a complex matrix $\bfR$:
\eqqn\eIIIxv{
\left[ \bfR \bfLambda \bfR^{-1} \right]_{\beta\alpha} = \lambda^2_\alpha 
\delta_{\beta\alpha}
}
where, in brief we obtain
\eqqn\eIIIxvi{
\eqalign{
\lambda^2_S = &
\frac{\displaystyle 1}{\displaystyle 2}\left[ 
\left( \Lambda_{11} + \Lambda_{22} \right) - Q \right] \cr
\lambda^2_L = &
\frac{\displaystyle 1}{\displaystyle 2}
\left[ 
\left( \Lambda_{11} + \Lambda_{22} \right) + Q \right] \cr}
}
with
\eqqn\eIIIxvii{
Q = \sqrt{
\left( \Lambda_{11} - \Lambda_{22} \right)^2 + 4 \Lambda_{12}\Lambda_{21} }
\; .}
The physical fields $\kl$ and $\ks$ corresponding to the 
eigenvalues
\eqqn\eIIIxviii{
\lambda^2_{S, L} = \left( m_{S, L} - 
\frac{\displaystyle i}{\displaystyle 2} \gamma_{S,L}
\right)^2 \simeq m^2_{S, L} - i m_{S,L} \gamma_{S,L}
}
are combinations of the $\ko$ and $\kob$
for which only the diagonal
elements of the propagator matrix contain poles in the $\kl$, $\ks$ basis
according to the following expression
\eqqn\eIIIn{
\Delta^{\prime}_{\alpha\beta} (q^2)
=
\left( \matrix{
{1 \over {q^2 - \lambda^2 _S}} & 0 \cr
0 & {1 \over {q^2 - \lambda^2 _L}} \cr } \right) \; .
}
In this approximation,
extracting the two non degenerate poles $\lambda^2_\alpha$,
we obtain the following form for the time evolution matrix
\eqqn\eIIevol{
\Uscr (t) = 
\left(
\matrix{
g_1 (t) &  \eta g_2 (t) \cr
{1\over\eta} g_2 (t) & g_1 (t) \cr}\right)
}
where $g_{1,2} (t) = {1\over 2} ( V_{SS} \pm V_{LL} )=
{1\over 2} (\hbox{e}^{- {\rm i} \lambda_{S} t} \pm
   \hbox{e}^{ - {\rm i} \lambda_{L} t} ) $.

\noindent
Nevertheless, since the LOY method is an approximate theory,
it is not surprising that it cannot satisfy exactly the unitarity 
requirement~\kapil ,
which is essential for the basic interpretation of any theory. The 
assumption of constant decay rates, as they arise by the unitarity sum rules
\BS \ connecting the kaon system with the space of all the decaying final 
states, then, cannot be justified in an exact sense.
In fact, for instance, the modulus of the ratio of the off-diagonal elements
\eqqn\eIIxiii{
r(t) ={ {U_{12}(t)} \over {U_{21}(t)} } = \left({ q\over p}\right)^2
}
differs from unity and could vary with time~\KAK \ 
if the time-reversal $T$--invariance is not a symmetry 
of the underlying Hamiltonian. In fact, in this case,
the condition of reciprocity is not properly satisfied~\kapil .
Indeed, the inclusion of off-diagonal terms
$V_{SL}=-V_{LS}$ in the 
evolution matrix $\Vscr$, induces a modification both of the time
evolution matrix and also of the previous ratio.
Assuming a global $CPT$--invariant propagation, in fact, it becomes
\eqqn\eIIxiiib{
r(t) = 
\left({ q\over p}\right)^2
\left({{1 - A}\over{1 + A}}\right) \quad.
}
On more general assumptions like causality and analycity,
the time dependence of the vacuum regeneration term 
$A=(V_{LS}-V_{SL})/(V_{SS}-V_{LL})$
can be obtained to study the $s=q^2$ dependence of the
proper self energy contributions $\bfPi$.
In particular, the very short time behaviour cannot be derived
taking solely into account the mentioned poles. The non exponential 
contributions can be estimated with somewhat greater generality 
to introduce some analytical techniques used in the recent literature on 
decay problems
\ref\insta{D. Cocolicchio, ``{\it The Characterization of Unstable Particles}",
preprint Sezione INFN Milano IFUM-FT 514/96}.
In fact, there is a resorted interest about the validity of the 
exponential decay law.
In general, the time evolution of a metastable 
quantum state is roughly described by three distinct trends. At very 
short time, the decay rate were noted to be characterized by a 
Gaussian behaviour. The exponential decay is expected within a 
limited time interval. An inverse power law will remain at long 
times in dependence of the structure of the initial state.
Technically, we can say that at short times with respect to the inverse width 
there is no exponential evolution because 
other poles, lying further from the real axis, may become important.
On the other side, for very long times, the cut contribution exceeds the exponential decrement
so that we get a residual inverse power law.
We shall not be concerned with the power law here, but 
we confine to study the breakdown of the exponential decay law due to the 
occurrence of bound states (poles on the real energy axis).
Therefore, we neglect the effects of the interactions of 
the (in)elastic rescattering of the final decay modes 
(branch points on the real axis).
However, the matrix elements of $\bfPi (s)$ have a cut along the positive real 
axis whose threshold is $m^2_{th}$ and are expected to have no poles 
there (Herglotz property). If the interaction is small, 
$\bfDelta (s)$ has two poles and we obtain the exponential law for 
the evolution of the kaon complex. But since the pole residues are 
nonorthogonal matrices rather than numbers, the dynamical semigroup
law Eq.~\law \  of the time evolution is violated, unless there is 
some symmetry which forces the Hamiltonian to be Hermitian.
Nevertheless, new poles emerging from the analytical continuation
become very important when they are close enough to the axis.
The dynamics for an eventual complex pole $s_p$ of the matrix propagator 
in the second Riemann sheet is however regulated by the following 
relation which locates the position of the poles
\eqqn\eIIxivb{
\det \Big[ s_p \delta_{\alpha\beta} - \bfLambda_{\alpha\beta} \Big] = 0 
\; , }
where $\bfPi$ and hence $\bfLambda$ are evaluated in the second 
Riemann sheet. Of course, these eventual poles become more important as they 
are closer to the real axis. The factorization of the $s_p$ pole in 
the transition propagator, yields
\eqqn\eIIxvb{
\bfDelta^{\prime -1} (s) = (s-s_p) {\pmb{\CZ}}^{-1} \; .}
Notice that we have absorbed
all renormalization effects into the matrix ${\pmb{\CZ}}$ which represents 
the residue of the full propagator at the $s_p$ pole. Extracting the 
leading term of the Laurent expansion about $s_p$, in terms of 
renormalized quantities we obtain 
\eqqn\eIIxvib{
{\pmb{\CZ}}^{-1} = [ \bfI - \bfPi^\prime (s_p) ] \; .}
However, it is possible to determine an exact expression for the residue
of the full propagator in terms of the projection operators
\eqqn\eIIIyvib{
\Pscr _{S, L} = \ket {K_{S, L} (s)}\bra{K^{\prime}_{S, L} (s)} \quad .}
In order to obtain an explicit form for ${\CZ}_ {\alpha, \beta}$, we can separate
the reduced resolvent $\bfDelta^{\prime}$ into two parts
\eqqn\eIIIIyvib{
\bfDelta^{\prime} (s) = {\Pscr _S (s) \over {s - \lambda^2_S (s)}} +
{\Pscr _L (s) \over {s - \lambda^2_L (s)}} }
where $\lambda^2_{S, L}(s)$ denote the two eigenvalues of $\bfLambda$
and $\Pscr _{S, L}$ satisfy the following properties
\eqqn\eIIIIyvib{
\cases{
\Pscr _S (s) \Pscr _L (s) =  \Pscr _L (s) \Pscr _S (s) = 0 \cr
 \cr
(\Pscr _{S, L})^2 =  \Pscr _{S, L} \quad . \cr}}
\blankline
Hence $\bfDelta^\prime$ has two simple poles corresponding to the values $s_S$ and
$s_L$, obtained by the equations
\eqqn\equat{
s_S - \lambda^2_S (s_S) =0, \quad s_L - \lambda^2_L (s_L) =0  \quad .}
Therefore for any given $s$, $\bfLambda (s)$ may be brought to a diagonal form by
a complex transformation whose columns are determinated by these eigenvalues. In 
the case that $\bfLambda$ may be treated as a constant, independent of $s$, over the entire 
range of interest, this transformation is independent of $s$.
A direct calculation of the residue of the propagator at the $s_S$
yields (an analogous expression is given for $\pmb \CZ (s_L)$ replacing
the subscript $L$ with $S$)
\eqqn\propagat{
\pmb \CZ (s_S) =
{\Pscr _S (s_S)
\over \phantom{s=}{1 - { d\lambda^2_S (s) \over ds}} \Big\vert _{s=s_S}}
\quad .  }
Of course, we could obtain a similar result from the viewpoint of
dispersion relations, using the separation of the one-particle states
from the $q^2$--dependent terms and indeed without talking about fields
at all~\insta . 

\blankline
\leftline {\bf III. Unitarity Sum Rules}
\blankline
\noindent     
The question of the correct treatment of the complex kaon faced with 
the problem that we must consider an Hilbert space $\Hscr = \Hscr_K 
\oplus \Hscr_F$, composed by the two dimensional neutral K--meson space 
$\Hscr_K$ and the space $\Hscr_F$ of all their decay final states. The 
evolution of the entire system should be defined by a semi--group of 
unitary operators in $\Hscr$ with a total self--adjoint Hamiltonian.
Therefore, once the decay channels being included into the base, the 
resulting total Hamiltonian will respect a semi--group evolution.
If we suppose that the set of final states $\{ F \}$ forms an 
orthonormal basis in $\Hscr_F$, the phenomenological description of 
the ${\rm K}^ 0 \overline{\rm K}{}^0$ meson decays
\ref\Peccei{S. H. Aronson, G. J. Bock, H.-Y. Cheng and E. Fishbach,
Phys. Rev. {\bf D28} (1983) 495;
L. Lavoura, Ann. Phys. {\bf 207} (1991) 428;
C. D. Buchanan et al., Phys. Rev. {\bf D45} (1992) 4088;
M. Hayakawa and A. I. Sanda, Phys. Rev. {\bf D48} (1993) 1150;
Zhi-zhong Xing, Phys. Rev. {\bf D53} (1996) 204}\ 
consists in the application of the LOY approximation supplemented by 
the unitarity sum rule of Bell and Steinberger~\BS \ connecting the 
two dimensional kaon subspace with the space of all final states for 
the decaying system.
In order to show the generalization of the Bell and Steinberger's 
relation, it is necessary to be specific about their assumptions.
The time evolution of the entire state vector $\ket{\bfPsi}$, composed 
of the neutral K-meson system with its decay channels $F$, is governed 
by the total Hamiltonian and it satisfies a completeness relation
\eqqn\compl{
\sum_i \vert \langle K_i \vert \bfPsi (t) \rangle\vert^2 + 
\sum_F \vert \langle F \vert \bfPsi (t) \rangle\vert^2 =
\vert\vert \bfPsi \vert\vert^2 \quad .
}
Introducing the reduced evolution $\Uscr$ in $\Hscr_K$ in contour representation
(we omit the prime index for brevity) 
\eqqn\brev{
{\cal U}_{ij}(t) = {1 \over 2{\pi}i}\int^{}_{{\rm Spec}(H)} ds \, e^{-ist} 
\bfDelta^{\prime}_{ij}(s)
}
where the integration extends over the whole spectrum of the Hamiltonian,
at any given time, initial kaon pure states (${\cal U}_{ij}(0) = \delta _{ij}$)
decay into an $F$ channel mode with an amplitude probability
\eqqn\amplit{
A_{iF}(t) = \left\langle {F} \vert {K_i (t)}\right\rangle =
\left\langle F \vert {\cal U}_{ij}(t) \vert K_j \right\rangle 
}
being 
\eqqn\bein{
A_D = (A_{iF}) = 
\left(\matrix{\left\langle {F} \vert {\ko}\right\rangle \cr
\cr
                   \left\langle {F} \vert {\kob}\right\rangle \cr}\right) =
\left(\matrix{A (\ko \rightarrow F) \cr
\cr
              A (\kob \rightarrow F) \cr}\right) \quad .
}
Squaring, we obtain the time--dependent rates
\eqqn\tdep{
\bfGamma ({\rm K}_i (t) \rightarrow F) = \sum^{}_{kj} {\cal U}^{\ast}_{
ki}{\cal U}_{kj} \vert A_{jF} \vert ^2
}
where the summation over repeated indices is tacitly assumed.
Then, the rate of decay into the final state $F$ at time $t$ can
be described in terms of the complex matrix elements
${\cal M}_{ij} = g_i g^\ast_j$ and of the amplitude probability at
time $t=0$ 
\eqqn\decakk{
\eqalign{
\bfGamma (\ko (t) \rightarrow F) = &
\vert A_F \vert^2 \Big[{\cal M}_{11}  + \vert \xi _F \vert ^2
{\cal M}_{22} - 2 {\rm Re} \Big( \xi_F {\cal M}_{21}\Big)\Big] \cr
\bfGamma (\kob (t) \rightarrow F) = &
\vert A_F \vert^2 \Big[{\cal M}_{22}  + \vert \xi _F \vert ^2
{\cal M}_{11} -2{\rm Re} \Big( \xi_F {\cal M}_{12} \Big)
\Big] \vert \eta \vert ^{-2} \quad . \cr}
}
Using the explicit form of ${\cal M}_{ij}$, the following general
formula for the time--dependent decay rates can be derived
\eqqn\dekayob{
\eqalign{
\bfGamma (\ko (t) \rightarrow F) =  
\vert A_F \vert ^2 e^{- \Gamma t} {1 \over 2}
\biggl\{ & \cosh \Big({\Delta \gamma \over 2} t \Big) +  \cos(\Delta m t)
+ \cr
+ & \vert \xi_F \vert^2 \Big[
\cosh \Big({\Delta \gamma \over 2} t \Big) - \cos(\Delta m t) \Big]+ \cr
+ & 2{\rm Re}\biggl[\xi_F \Big[\sinh \Big({\Delta \gamma \over 2}t \Big)
+ i \sin (\Delta m t ) \Big] \biggr] \biggr\}\cr}}
for $\ko (t)$ and
\eqqn\koddecay{
\eqalign{
\bfGamma (\kob (t) \rightarrow F) =
\vert A_F \vert ^2 e^{- \Gamma t} {1 \over 2}
\biggl\{ & \cosh \Big({\Delta \gamma \over 2} t \Big) -  \cos(\Delta m t)
+ \cr
+ & \vert \xi_F \vert^2 \Big[
\cosh \Big({\Delta \gamma \over 2} t \Big) +  \cos(\Delta m t) \Big]+ \cr
+ & 2 {\rm Re}\biggl[\xi_F \Big[\sinh \Big({\Delta \gamma \over 2}t \Big)
-i \sin (\Delta m t ) \Big] \biggr] \biggr\}
\vert \eta \vert ^{-2}\cr}}
for $\kob (t)$.
For convenience, in the previous formulas we have introduced
the complex parameter
$\xi_F={\displaystyle q \over \displaystyle p}
{{\displaystyle A  (\kob\rightarrow F)} \over
{\displaystyle A (\ko\rightarrow F)}}$ and the quantities
\eqqn\quant{
\matrix{
\Delta m = m_S -m_L & m={{\displaystyle m_S + m_L} \over \displaystyle 2}
                                            \cr
                    &                       \cr
\Delta \gamma = \gamma _S - \gamma _L & \Gamma = {{\displaystyle \gamma_S +
\gamma_L} \over \displaystyle 2}  \quad . \cr}
}
The time--integrated rates of interest
\eqqn\widhts{
{\widehat {\bfGamma}}({\rm K}_i \rightarrow F) = 
\int_0^\infty dt \, \bfGamma({\rm K}_i (t) \rightarrow F)=
\int_0^\infty dt \,
\vert A_{iF} (t) \vert ^2 \quad ,
}
which with the position ${\widehat{\cal M}}_{ij} = \int_0^\infty dt \, 
{\cal M}_{ij} (t)$ become 
\eqqn\becom{
\eqalign{
 &{\widehat {\bfGamma}}(\ko \rightarrow F) = 
\vert A_F \vert^2 \Big[{\widehat{\cal M}}_{11}  + \vert \xi _F \vert ^2
{\widehat{\cal M}}_{22} - 2 {\rm Re} \Big( \xi_F {\widehat{\cal M}}
_{21}\Big)\Big] \cr
& {\widehat {\bfGamma}}(\kob \rightarrow F) = 
\vert A_F \vert^2 \Big[{\widehat{\cal M}}_{22}  + \vert \xi _F \vert ^2
{\widehat{\cal M}}_{11} - 2 {\rm Re} \Big( \xi_F {\widehat{\cal M}}
_{12}\Big)\Big]
\vert \eta \vert ^{-2} \quad ,
\cr }
}
let us write the content of the Bell and
Steinberger's unitarity sum rule:
\eqqn\SRBS{
\eqalign{
\langle K_S \vert \bfGamma \vert K_L \rangle = &
\Big[ \big( {{\gamma_S +\gamma_L}\over 2} \big) - i \big( 
m_S-m_L\big)\Big] \left\langle {K_S} \vert {K_L}\right\rangle = \cr
= & \sum_F \langle F\vert H_K \vert K_S\rangle^\ast 
\langle F\vert H_K\vert K_L\rangle = 
\sum_F \langle \Gamma_F\rangle \big( \Ascr + i \Bscr) \cr}
}
where the last expression has been obtained by choosing the final 
decay modes $F$ to be CP-eigenstates and the integration
with respect to the phase space must be understood. We have defined
\eqqn\Fgam{
\langle \Gamma_F\rangle  =
{1 \over 1+\vert \eta \vert ^2} \Big[ \bfGamma(\ko\rightarrow F)
+  \vert \eta \vert ^2 \bfGamma(\kob\rightarrow F) \Big] \simeq
{1\over 2}
\Big[ \bfGamma(\ko\rightarrow F) +
\bfGamma(\kob\rightarrow F)\Big]
\; .}
The two independent CP--violating parameters are given by
\eqqn\AeB{
\eqalign{
\Ascr=& {{1-\abs{\xi_F}^2} \over {1+\abs{\xi_F}^2} } \cr
\Bscr=& {{2\Im\xi_F}\over{1+\abs{\xi_F}^2} }\cr}
}
and
$\langle {\rm K}_L\vert {\rm K}_S \rangle
={ { 1- \vert \eta \vert ^2 } \over {1+
\vert \eta \vert ^2 }}\simeq 10^{-3}$
imposes large cancellations in the sum.
The first of these parameters ($\Ascr$) vanishes in the absence of 
$\kok$ mixing. By contrast, the second parameter ($\Bscr$) yields
$CP$--violation in the decay amplitude (i.e. $\abs{\xi_F}\neq 1$).

\noindent
We can reformulate this unitarity sum rule by means of the 
propagator method. The propagator appears in the scattering amplitude 
as a factor sandwiched between vertex functions referring to the 
particular processes in which the unstable system is produced and 
detected. The vertex functions of the produced (P) and decay (D) 
positions are assumed to be slowly varying functions of momentum over 
the range corresponding to the width of the mesons.
Evidently, until the dynamics of the meson system has been described 
in terms of initial and final asymptotic states, it results unitary 
and causal. Actually, the decay rates of the neutral kaons are so 
small to legitimate the use of $\ko$ and $\kob$ as two   
asymptotic states of the scattering S-matrix.
Of course, this view is not justified for a very short time interval 
(much shorter than the mean life of $K_S$). In fact in this time 
interval, decay processes cannot be described only by the pole 
dynamics. 

\noindent
Consider a process of the sort studied in CPLEAR experiments and 
schematically shown in Fig.~1. For definiteness, let us assume that 
the external lines attached to the production vertex P represent the 
initial I channel mode (for example $p {\bar p}$), the internal line 
represents the neutral kaon ${\rm K}^ 0 \overline{\rm K}{}^0$ 
complex and the external lines 
attached to the decay vertex D represents the final F decay mode (for 
example two pions). At the CPLEAR experiment, $\ko$ and $\overline{\rm K}{}^0 $ are 
produced at point $x$ in the strong interaction of $p{\bar p}$ and 
subsequently decay at point $x^\prime$ to $\pi^+\pi^-$.
In pole approximation,
the description of the time evolution assumes that the $q^2$ dependence 
of the matrix elements comes from the intermediate propagators, and 
none from the vertices. As a first approach,
it does not seem an excessively
drastic truncation to use the pole approximation and to neglect the 
dependence of $\bfPi$ on $q^2$ in a large range. This pole 
approximation consists in replacing the full propagator with the pole 
form of Eq.~\eIIIn \ and the two vertices by their values at the pole.
First of all, we want now to show the connection of this formalism with the 
usual properties of particle mixing in which we solve a Schr\"odinger 
equation for an effective Hamiltonian ${\pmb{H}}_K={\pmb{M}}- {i\over 2}
\bfGamma$ acting on a Hilbert space in which the only states are 
one-particle states. It is characteristic of such approximation that
the eigenvalues and the eigenstates of the effective Hamiltonian are 
found from the columns of the diagonalizing complex matrix $\bfR$. The 
effective Hamiltonian is not Hermitean thus these eigenstates are not 
in general orthogonal because of the existence of common decay 
channels. In comparing these methods, note that in the propagator 
method the notation refers to the square of an effective complex mass 
matrix
\eqqn\hamil{
\bfLambda\simeq {\pmb{M^2}} - i {\pmb{M}}\bfGamma 
}
whereas the effective Hamiltonian $H$ is decomposed according to the 
mass and decay Hermitean matrices
\eqqn\MandG{
{\pmb{M}} = {1\over 2}\Big( H+ H^{\dag}\Big) \quad {\rm and} \quad
 \bfGamma = i\Big( H - H^{\dag} \Big)  
}
In the single pole approximation, the two formalisms become equivalent 
if we neglect terms of order $\bfGamma/4$, so that we approximate
\eqqn\lapr{
\lambda_{S,L}^2\simeq m_{S,L}^2 - i m_{S,L}\gamma_{S,L}\; .
}
As we stressed above, within a sensible energy interval, the 
production ${\widehat {\pmb{A}}_P} = ({\widehat A}_{iI})$ and decay 
${\widehat {\pmb{A}}}_D=({\widehat A}_{jF})$ 
amplitudes can be considered as momentum--independent quantities.
The general s-channel scattering amplitude can be extracted from the 
diagrams of Fig.~2. The 
transition amplitude is given in a matrix notation by
\eqqn\btran{
{\Ttr}_{FI} = \sum_{ij} \hA_{Fi}^\ast\big[ s\bfI -\bfLambda\big]
_{ij}^{-1} \hA_{jI}=
{\widehat {\pmb{A}}}_D^{\dag}\big[ s\bfI -\bfLambda\big]^{-1}
{\widehat {\pmb{A}}}_P
}
where ${\widehat {\pmb{A}}_{D,P}}$ represent the transition amplitudes 
in the propagator notation.
After the diagonalization of $\bfLambda$, in 
terms of the physical states
the transition matrix takes the form
\eqqn\takes{
\Ttr_{FI} = \sum^{}_{\alpha} \biggl[{{\widetilde A}_{F \alpha}
{\widetilde A}_{\alpha I} \over { s-\lambda^2_\alpha}} \biggr] =
{{\widetilde A}_{FS} {\widetilde A}_{SI} \over { s-\lambda^2_S}} +  
{{\widetilde A}_{FL} {\widetilde A}_{LI}
\over { s-\lambda^2_L}}
}
where the summation over the repeated indices and integration
over the phase space of the final states $F$ must be
understood. The unitarized tilded amplitude
${\widetilde {\bf A}} = {\bf R} {\widehat {\bf A}}$
represents the decay amplitude of a ${\rm K}_ \alpha$ eigenstate
into a decay channel $F$.
Due to the unitarity of the scattering matrix $S_{FI} =
\delta_{FI} -i \Ttr_{FI}$ and using the optical theorem
at the level of the effective anti Hermitian part of $\bfLambda$
in the particular case of real values of the Mandelstam $s$ variable,
we obtain
\eqqn\obtain{
{\bf R}\Big(\bfLambda -\bfLambda^{\dag}
\Big)_{\alpha \beta}{\bf R}^{\dag}  =
-i{\widetilde {\bf A}}_D {\widetilde {\bf A}}^{\dag}_D
= -i \sum^{}_F {\widetilde
A}_{\alpha F} {\widetilde A}^{\ast}_{\beta F} 
}
where
\eqqn\ieei{
\sum^{}_F {\widetilde A}_{\alpha F} {\widetilde A}^{\ast}_
{\beta F} = i (\lambda^2_\alpha - \lambda^{\ast 2}_\beta) {\cal O}_{
\alpha \beta}
}
where ${\bf {\cal O}}_{\alpha \beta} = ( {\bf R}{\bf R}^{\dag})_{\alpha \beta} = 
\left\langle {K_\alpha} \vert {K_\beta}\right\rangle $.
Eq.~\obtain\ and Eq.~\ieei\ represent nothing but the S--matrix version
of the Bell and Steinberger's relation. 
The total decay width of the state $\ket {K_\alpha}$ into the
decay channel $F$ is therefore
\eqqn\chann{
\sum^{}_F \bigl\vert {\widetilde A}_{\alpha F} \bigr\vert ^2 =
i(\lambda^2_\alpha - \lambda^{\ast 2}_\alpha)  \quad .
}
Although, we consider only poles corresponding to the mass and
the decay width of $\ks$ and $\kl$, one may consider the momentum
dependence of the effective complex $\bfLambda (q^2)$. In
the narrow width approximation the matrix elements of
$\bfLambda (q^2)$ are assumed to be constant in a wide energy
region. But we cannot use this approximation in all our
applications since in same energy region a substructural effect could
emerge. If the transition elements are meromorphic functions
in the complex momentum plane having poles in the points satisfying
equation ${\rm det}[s {\bf I} - \bfLambda]=0$, residues
in the poles are complex matrices. Let us introduce the resolvent
$\bfDelta ^{\prime} (s) = [s {\bf I} -\bfLambda]^{-1}$ and
we consider the transition matrix in the decay space
\eqqn\spacedd{
\Ttr(s) = {\widehat {\bf A}}^{\dag}_D {\bfDelta}^{\prime}(s) 
{\widehat {\bf A}}_P
}
similar to the Wigner ${\bf \Re}$--matrix of the nuclear reaction
theory. Actually, we neglect the non--resonant parts of the
transition amplitude, that is equivalent to consider isolated
narrow particles rather two overlapping resonances.
Instead of working this way, we can replace back from the
beginning the flavour propagator with the introduction of
renormalized matrix propagator Eq.~\eIIxvib.

\blankline
\leftline {\bf IV. Concluding Remarks}
\blankline
\noindent     
Usually, the Wigner--Weisskopf narrow width approximation has been 
applied to describe the law of the time evolution in weak nonleptonic 
decays of the $K^0$ and $B^0$ mesons, where some signals of 
$CP$--violation can be found experimentally. This technique is not
rigorous, in the sense that a relativistic formulation is required to 
deal with the unstable kaon system. Without assuming a 
specific dynamical model, we have analyzed the covariant propagator 
method. In the pole approximation, within the spirit of the mass--mixing 
formalism, we take the widths to be constant, with no explicit functional 
$q^2$--dependence.
This approach reproduces the usual 
exponential decay law of the physical kaon evolution. In this paper, 
we have shown the connection of the propagator formalism with the usual LOY 
approximation, in which we solve a Schr\"odinger equation for an 
effective Hamiltonian acting on a Hilbert space in which the only 
states are one--particle states.
In this effective approach,
working with mass rather than mass--squared matrix, no
momentum dependence results.
Indeed, we now consider the further contributions of
eventual other poles which can induce some relevant deviations.
They lie on the second Riemann complex momentum
sheet of the analytically continued matrix propagator closer to the 
cut along the real axis. We provide some phenomenological implications of 
these analytical complexities. Moreover, the $q^2$-- dependence of the 
propagator matrix could be incompatible with the constraints of 
unitarity and causality. Independently from the specific dynamics of a
given theory and from the subtleties of the higher--order terms in the matrix
elements of the propagator, which are relevant only close to the 
${\rm K}^ 0 \overline{\rm K}{}^0$ resonance, we propose a generalization
of the unitarity sum rules. All methods have advantages and disadvantages.
The standard time--dependent effective Hamiltonian formalism \Peccei\
is expressed directly in terms of the measured quantities and is therefore
more useful in a wide range of phenomenological applications.
However, the propagator method arises naturally in the context of
quantum field
theory and hence close to the requirements of the fundamental
gauge theories. Furthermore, as we have shown, this last method has
the considerable
advantage to provide a description of the additional complications
connected with
the $q^2$--dependence of the underlying interactions. In modern gauge
theories,
CP violating--asymmetries are generated from the interference of the
tree--level amplitude
with higher--order corrections to vertex, mass and wave--function.
CP violating--asymmetries occur through interference between at least two 
possible amplitudes having different weak and strong phases. In the
Standard Model
a weak--phase difference is provided by the different complex phase
of the CKM
matrix elements of the tree and penguin diagrams, while the strong
phase is given
by the absorptive parts of the corresponding diagram.
In this paper we analyzed carefully CP--violating effects
due to intermediate oscillating mesons. These effects can be established even for 
vanishing strong phases difference and could be relevant in models
of spontaneous CP--violation due to Higgs--bosons exchanges.

\vfil\eject
\vfill\eject\immediate\closeout\rfile
\centerline{{\bf References}}\bigskip
\input refs.tmp\vfill\eject
 \bye